\documentclass[prb,aps]{revtex4}
\usepackage[dvips]{graphicx}
\usepackage{amssymb}
\begin{document}


\title{Sharp bends in photonic crystal waveguides as nonlinear Fano resonators}

\author{Andrey E. Miroshnichenko and Yuri S. Kivshar}
\affiliation{Nonlinear Physics Centre and Centre for Ultra-high
bandwidth Devices for Optical Systems (CUDOS), Research School of
Physical Sciences and Engineering, Australian National University,
Canberra ACT 0200, Australia}

\begin{abstract}
We demonstrate that high transmission through sharp bends in
photonic crystal waveguides can be described by a simple model of
the Fano resonance where the waveguide bend plays a role of a
specific localized defect. We derive effective discrete equations
for two types of the waveguide bends in two-dimensional photonic
crystals and obtain exact analytical solutions for the resonant
transmission and reflection. This approach allows us to get a
deeper insight into the physics of resonant transmission, and it
is also useful for the study and design of power-dependent
transmission through the waveguide bends with embedded nonlinear
defects.
\end{abstract}

\maketitle

\section{Introduction}

Photonic crystals (PCs) are artificial dielectric structures with
a periodic modulation in the refractive index that create a range
of forbidden frequencies called a photonic
band gap~\cite{jdjprvsf97}. The existence of the photonic band gap
can change dramatically the properties of the light allowing the
realization of ultra-compact and multi-functional optical devices.
One of the most fascinating properties of photonic crystals is
their ability to guide electromagnetic waves in the narrow
waveguides created by a sequence of line defects, including light
propagation through extremely sharp waveguide bends with nearly
perfect power
transmission~\cite{amjcciksfprvjdj96,sylecvhprvjdj98}. It is
believed that the low-power reflection of sharp waveguide bends in
photonic crystals is one of the most promising approaches to
combine several devices inside a compact nanoscale optical chip.

The main advantage in achieving low radiation losses for the light
transmission through sharp waveguide bends is based on the
existence of the photonic band gap allowing to confine light inside
a narrow defect waveguide due to the effect of the resonant Bragg
scattering, reducing only the reflection losses of the bend
design. Recent studies addressed the issue of an improved design
of the sharp waveguide bends in two-dimensional photonic crystals
and suggested that the transmission losses can be less then $5\%$
\cite{amjcciksfprvjdj96,acsn00,rleruafpmjsrmoj01,ecsyljrwsgjjdj01,
sohbcwcjmstfkrhuo02,jschde03,zylkmh03,atmacmsmmsaphl04,jsjos04,inpprmdlr04,nm04}.
However, most of those studies are based on direct numerical
simulations and neither provide a deep understanding of the
physics of the enhanced transmission of the sharp waveguide bends
nor suggest simple approaches to design the waveguide bends with
required transmission properties.

The main purpose of this paper is twofold. First, we demonstrate
that transmission of electromagnetic waves through sharp bends of
photonic crystal waveguides can be described by a simple discrete
model where the waveguide bend plays a role of a specific
localized defect. By employing the semi-analytical approach based
on the Green's function
formalism~\cite{sfmyskras00,sfmysk02,sfmusk02}, we derive
effective discrete equations which allow us to describe the bend
transmission as a special case of the so-called Fano
resonance~\cite{uf61}, recently discussed for the wave propagation
in discrete chains~\cite{aemsfmsfyusk05}. We demonstrate how to
introduce the effective discrete model for the Fano resonance by
selecting two common designs of the waveguide bends in
two-dimensional photonic crystals created by a lattice of rods,
and obtain {\em exact analytical solutions} for the resonant
transmission and reflection. We show that asymmetric shapes of the
transmission curves observed for the waveguide bends  can be
understood in terms of the Fano resonance which originates from
the interaction between continuum waves and an effective localized
state associated with the waveguide bend that provides an
additional propagation channel for the wave transmission and,
therefore, leads to the constructive or destructive interference.
Second, we show that this approach allows us not only get a deeper
physical insight but it is also useful to study the nonlinear
transmission through the waveguide bends with nonlinear defects.

The paper is organized as follow. In Section 2 we derive the
effective discrete equations for the case of two-dimensional
photonic crystal waveguides based on the Green's function
technique. We employ the approach earlier suggested by Mingaleev
{\em et al.}~\cite{sfmyskras00,sfmysk02,sfmusk02}, but take into
account the effect of the long-range interaction only in the
vicinity of the waveguide bend. This allows us to derive an
effective discrete Fano model where the waveguide bend plays a
role of a special localized defect, and also suggest an effective
way for predicting and controlling the properties of different
types of the waveguide bends. Using this model, in Section 3 and
Section 4 we study the wave transmission in the linear and
nonlinear regimes, respectively. Section 5 concludes the paper.

\begin{figure}
    \centering
        \includegraphics[width=1.00\columnwidth]{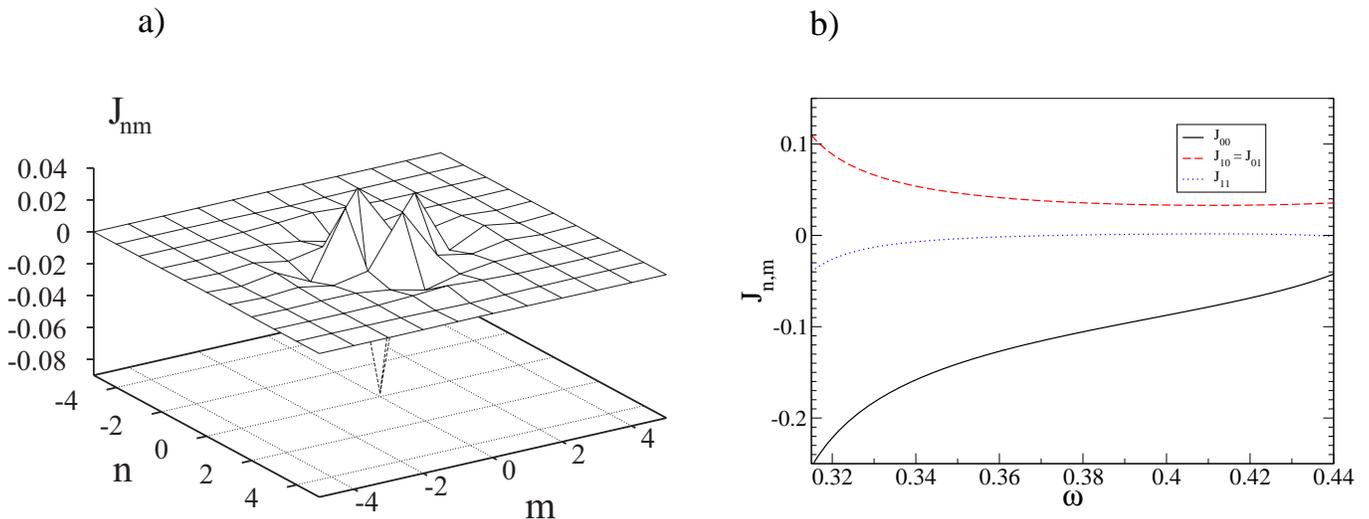}
\caption{ (a) Spatial structure of the coupling coefficients
$J_{nm}(\omega)$ of the effective discrete model
(\ref{discrete_model}) at $\omega = 0.4\times2\pi c/a$.
(b) Dependence of the specific coupling coefficients (marked) on
the frequency $\omega$.
     }
    \label{fig:fig1}
\end{figure}

\section{Effective discrete model}

First, we derive an effective discrete model for the wave
transmission through a sharp waveguide bend by employing and
modifying the conceptual approach suggested earlier by Mingaleev
{\em et al.}~\cite{sfmyskras00,sfmysk02}. We consider a
two-dimensional photonic crystal created by a square lattice (with
the period $a$) of dielectric rods in air ($\epsilon_{\rm bg}=1$).
The rods have the radius $r=0.18a$ and the dielectric constant
$\epsilon_{\rm rod}=11.56$. We study in-plane light propagation in
this photonic lattice described by the electric field
$E(\mathbf{x},t)=\exp(-i\omega t)E(\mathbf{x}|\omega)$ polarized
parallel to the rods, and reduce the Maxwell's equations to the
scalar eigenvalue problem
\begin{eqnarray}
\label{electric_field}
\left[\nabla^2+\left( \frac{\omega}{c}\right)^2\epsilon(\mathbf{x})\right]E(\mathbf{x}|\omega)=0\;.
\end{eqnarray}
For given parameters, this square lattice of rods is known to
possess a large TM band gap ($38\%$) between the frequencies
$\omega=0.303\times2\pi c/a$ and $\omega=0.444\times2\pi c/a$.

We create a waveguide by replacing some of the lattice rods by the
defect rods with the radius $r_d$, or simply by removing some rods
of the lattice. To describe the structure with defects, we
decompose the permittivity function $\epsilon(\mathbf{x})$ into a
sum of the periodic part and the defect-induced contribution,
$\epsilon(\mathbf{x})=\epsilon_p(\mathbf{x})+\delta\epsilon(\mathbf{x})$,
and rewrite Eq.~(\ref{electric_field}) in the integral
form~\cite{sfmyskras00},
\begin{eqnarray}
\label{integral}
E(\mathbf{x}|\omega)=\left( \frac{\omega}{c}\right)^2\int d^2\mathbf{y}
G(\mathbf{x,y}|\omega)\delta\epsilon(\mathbf{y})E(\mathbf{y}|\omega)\;,
\end{eqnarray}
where $G(\mathbf{x,y}|\omega)$ is the standard Green's function.
If the radius of the defect rod $r_d$ is sufficiently small, the
electric field $E(\mathbf{x}|\omega)$ inside the rod is almost
constant, and the integral (\ref{integral}) can be easily
evaluated. This allows us to derive a set of discrete equations
for the electric field
\begin{eqnarray}
\label{discrete_model}
E_{n,m}=\sum\limits_{k,l}J_{n-k,m-l}(\omega)\delta\epsilon_{k,l}E_{k,l}\;
\end{eqnarray}
where
\begin{eqnarray}
J_{n,m}(\omega)=\left(\frac{\omega}{c}\right)^2\int\limits_{r_d} d^2\mathbf{y}
G(\mathbf{x}_n,\mathbf{x}_m+\mathbf{y}|\omega)
\end{eqnarray}
are the frequency-dependent effective coupling coefficients,
\begin{eqnarray}
\delta\epsilon_{n,m}=\epsilon_{n,m}-\epsilon_{\rm rod}\;,
\end{eqnarray}
is the defect-induced change of the lattice dielectric function,
where $\epsilon_{n,m}$ is the dielectric constant of the defect
rod located at the site $(n,m)$.

\begin{figure}
    \centering
        \includegraphics[width=.8\columnwidth]{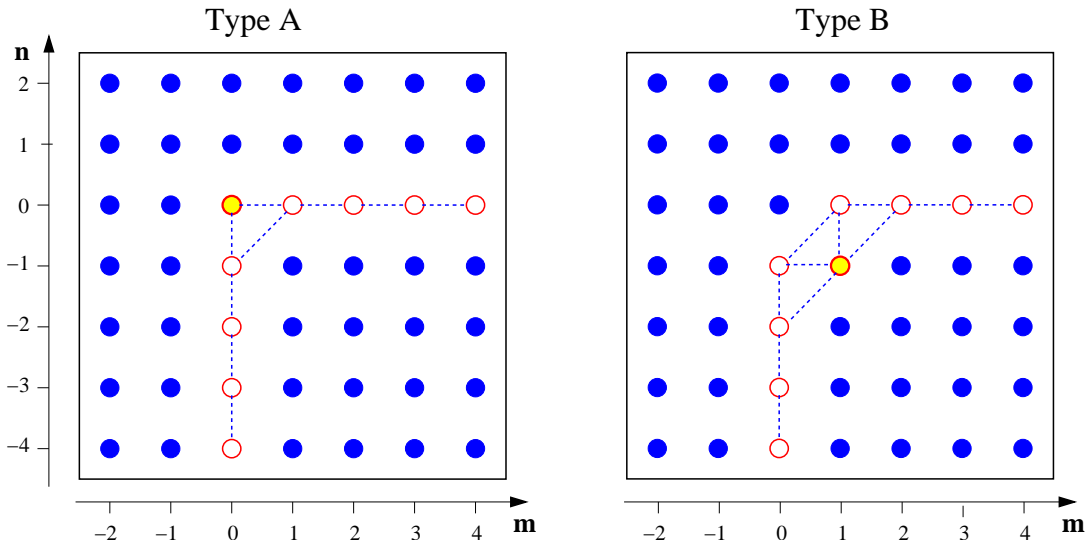}
    \caption{
Schematic view of two designs of the waveguide bends studied in
the paper. Empty circles correspond to removed rods, dashed lines
denote the effective coupling. Yellow circles mark the defects with
different dielectric constant $\epsilon_d(|E|)$, which can be
nonlinear. }
    \label{fig:fig2}
\end{figure}

In general, the effective coupling coefficients
$|J_{n,m}(\omega)|$ decay slow in space, as shown in
Fig.~\ref{fig:fig1}(a). This slow decay introduces effective {\em
long-range interaction} (LRI) from site to site of the waveguide,
which becomes crucially important for the waveguide bends. In
reality, we define a finite distance $L$ of this (formally
infinite) interaction by assuming that all coupling coefficients
with the numbers $|n-k|>L$ and $|m-l|>L$ vanish. As demonstrated
in Ref.~\cite{sfmysk02}, the case $L=6$ gives already an excellent
agreement with the results of the finite-difference time-domain
numerical simulations.

Unlike the previous studies of the effective discrete model, here
we are interested in the light propagation through two types of
the $90^0$ waveguide bends shown in Fig.~\ref{fig:fig2} created by
removed rods and additional defect rods with dielectric constant
$\epsilon_d$ placed at the corner; the defect rod can also be
nonlinear. Due to the effective long-range interaction, the corner
in the bend waveguide can be considered as a special type of a
defect; this resemblance is indeed correct and it explains the
Fano-type resonances in the transmission as discussed below.

\section{Linear transmission}

In this section, we consider the linear transmission through the
waveguide bends when all corner defects are assumed to be linear.
In order to show analytically that the waveguide bend generates an
effective defect state and it can be described by a discrete model
of the Fano resonance, we proceed in two steps making
simplifications in the general model (\ref{discrete_model}). For
simplicity, we consider the waveguide bend of the type A, as shown
in Fig.~\ref{fig:fig2}.

As the first step, we reduce the effective length of the
long-range interaction in the straight parts of the waveguide bend
taking it as $L=1$. This approximation corresponds to the familiar
tight-binding approximation that takes into account only the
coupling between the nearest neighbors in a discrete chain.
However, we take into account the nonlocal coupling near the sharp
bend, and this requires to introduce an additional coupling of
two sites around the corner as shown in Fig.~\ref{fig:fig2} for
the bend of the type A. It turns out that this simple
approximation allows us to describe qualitatively all important
properties and the key physics of the waveguide bend.

To derive the effective discrete model corresponding to the bend
design shown in Fig.~\ref{fig:fig2} (left), we remove two
semi-infinite rows of rods and write the dielectric function in
the form,
\begin{eqnarray}
\epsilon_{n,m} = \epsilon_{\rm bg}\;,\;\; \mathrm{for}\;\;
n\le-1\; \mathrm{and}\; m=0\;\; \mathrm{or}\;\; n=0
\;\mathrm{and}\; m\ge1\;.
\end{eqnarray}
At the corner of the waveguide bend, we place a defect rod with
the dielectric constant $\epsilon_{0,0}=\epsilon_d$. Equations
(\ref{discrete_model}) for the electric field inside the defect
rods  can be written explicitly as
\begin{eqnarray}
\label{tbmodel}
(1-J_{0,0}\delta\epsilon_0)E_{n,0} &=&\delta\epsilon_0J_{1,0}(E_{n+1,0}+E_{n-1,0}),\;\;n<-1,\nonumber\\
(1-J_{0,0}\delta\epsilon_0)E_{0,m} &=&\delta\epsilon_0J_{0,1}(E_{0,m+1}+E_{m,n-1}),\;\;m>1,\nonumber\\
(1-J_{0,0}\delta\epsilon_0)E_{-1,0} &=&J_{1,0}(\delta\epsilon_1E_{0,0}+\delta\epsilon_0E_{-2,0})+
J_{1,1}\delta\epsilon_0E_{0,1},\;\;\\
(1-J_{0,0}\delta\epsilon_1)E_{0,0} &=&\delta\epsilon_0(J_{0,1}E_{0,1}+J_{1,0}E_{-1,0}),\;\;\nonumber\\
(1-J_{0,0}\delta\epsilon_0)E_{0,1}
&=&J_{0,1}(\delta\epsilon_1E_{0,0}+\delta\epsilon_0E_{0,2})+
J_{1,1}\delta\epsilon_0E_{-1,0},\;\;\nonumber
\end{eqnarray}
where $\delta\epsilon_0=\epsilon_{\rm bg}-\epsilon_{\rm rod}$ and
$\delta\epsilon_1=\epsilon_{d}-\epsilon_{\rm rod}$. Importantly,
the resulting set of coupled equations present a discrete model
that can be compared with the discrete models earlier studied in
Ref.~\cite{aemsfmsfyusk05}, for which the existence of the Fano
resonance has been demonstrated analytically.

First two equations in the system (\ref{tbmodel}) allow us to
obtain the dispersion relation for the bend waveguide far away
from the corner. Due to the symmetry of the photonic crystal, the
coupling terms $J_{0,1}$ and $J_{1,0}$ coincide and, therefore, we
obtain the waveguide dispersion in the form
\begin{eqnarray}
\label{dispersion} \cos
k=\frac{1-\delta\epsilon_0J_{0,0}}{\delta\epsilon_0J_{0,1}},
\end{eqnarray}
where $k$ is the wavenumber for the waves propagating along the
waveguide. Other equations in the system (\ref{tbmodel}) allow us
to calculate the transmission coefficient of the waveguide bend,
\begin{eqnarray}
\label{transm}
T=\frac{4a^2\sin^2k}{|b(c-b)|^2}\;,
\end{eqnarray}
where we use the notations
\begin{eqnarray}
\label{coeff}
a=(J_{11}+J_{0,1}^2\delta\epsilon_1-J_{0,0}J_{1,1}\delta\epsilon_1)J_{0,1}\delta\epsilon_0^2\;,\;\;
b=(J_{0,0}+ \exp(ik)J_{0,1}-J_{1,1})\delta\epsilon_0-1\;,\nonumber\\
c=(J_{0,0}^2-2J_{0,1}^2+J_{0,0}J_{1,1}+
\exp(ik)J_{0,0}J_{0,1})\delta\epsilon_0\delta\epsilon_1-J_{0,0}\delta\epsilon_1\;.\;\;\;\;
\end{eqnarray}
Important information about the resonant transmission follows from
the study of zeros of the transmission coefficient (\ref{transm}).
Except for the band edges ($k=0,\pi$), the transmission
coefficient vanishes when $a=0$, i.e. when
\begin{eqnarray}
\label{zeros}
J_{0,0}J_{1,1}\delta\epsilon_1=J_{11}+J_{0,1}^2\delta\epsilon_1.
\end{eqnarray}

\begin{figure}
    \centering
        \includegraphics[width=.8\columnwidth]{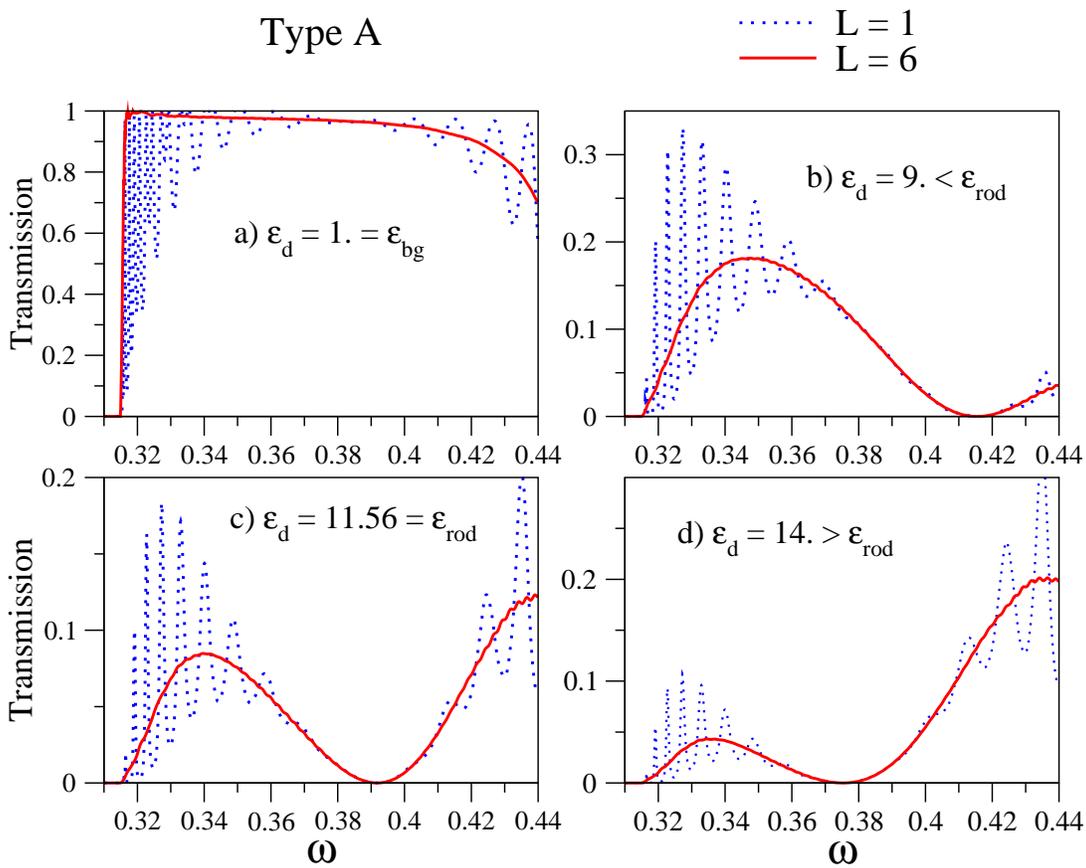}
    \caption{
Transmission coefficient of the waveguide bend for different
values of the dielectric constant $\epsilon_d$. The Fano resonance
is observed when the value of the dielectric constant of the
defect rod $\epsilon_d$ approaches the value of the dielectric
constant of the lattice rod $\epsilon_{\rm rod}$. The plot
$\epsilon_d=\epsilon_{\rm bg}$ corresponds to the case when a rod
is removed from the bend corner, whereas the plot
$\epsilon_{d}=\epsilon_{\rm rod}$ corresponds to the case when the
lattice rod remains at the corner.
     }
    \label{fig:fig3}
\end{figure}

As the second step, we approximate the dependence of the coupling
coefficients $J_{n,m}$ in the effective discrete model on the
frequency $\omega$, considering the specific results presented in
Fig.~\ref{fig:fig1}(b). In the frequency interval $[0.32,0.42]$,
we find the following approximation
\begin{equation}
\label{approx}
    J_{0,0}=\omega-\omega_d, \;\;\;\; J_{0,1}=C, \;\;\;\;
J_{1,1}=VJ_{0,0}=V(\omega-\omega_d),
\end{equation}
where $\omega_d$ is the frequency of a single embedded defect (in
our case, $\omega_d \approx 0.4$), $C$ and $V$ are constants
($C\approx0.035$ and $V\approx0.07$). By substituting
Eq.~(\ref{approx}) into Eq.~(\ref{zeros}), we obtain the quadratic
equation
\begin{eqnarray}
\label{quadratic}
\delta\epsilon_1V\omega^2-(2\delta\epsilon_1\omega_d+1)V\omega+
(V\delta\epsilon_1\omega_d^2+V\omega_d-C^2\delta\epsilon_1)=0,
\end{eqnarray}
which has two solutions
\begin{eqnarray}
\omega_F=\omega_d+\frac{1}{2\delta\epsilon_1}\pm\left[4C^2+\frac{V}{4\delta\epsilon_1^2}\right]^{1/2}.
\end{eqnarray}

According to these results, there exists a possibility for two
perfect reflections and, therefore, two Fano resonances. This is
due to the linear frequency dependence of two coupling terms
(\ref{approx}), the on-site term $J_{0,0}$ and the cross-coupling
term $J_{1,1}$. In our model, we have only one free parameter---
the dielectric constant $\epsilon_d$ of the defect rod at the
corner. Then, when $\delta\epsilon_1 \approx 0$ at list one zero
of the transmission lies inside the propagation spectrum
(\ref{dispersion}), $\omega_0<\omega_F<\omega_{\pi}$. As a result,
we predict that when $\epsilon_d\approx\epsilon_{\rm rod}$ the
perfect reflection through the waveguide bend of the type A should
be observed, and numerical results confirm this prediction (see
Fig.~\ref{fig:fig3}). We plot the transmission coefficient for two
different values of the interaction, $L=1$ and $L=6$. This shows
that the tight-binding approximation ($L=1$) still gives
reasonable results in average, and it works especially well near
the Fano resonance where our model works nicely.

Both the effective model and calculation of the transmission
coefficient for the waveguide bend of the type B (see
Fig.\ref{fig:fig2}) are similar. The only difference between these
two types of the bend design is the existence of additional
coupling terms to the the defect state. As was shown in
Ref.~\cite{aemsfmsfyusk05}, by increasing the number of the
coupling terms we only shift the position of the Fano resonance,
either transmission or reflection. In our case, the coupling is
small ($V\ll1$), so that the renormalization is negligible. For
the waveguide bend of the type A,  the Fano resonance manifests
itself as the perfect reflection only, whereas the perfect
transmission lies outside the waveguide spectrum. By analyzing the
transmission coefficient for the type B bend, we can show that in
this case the perfect transmission may move to the waveguide
spectrum when $\epsilon_d\approx\epsilon_{\rm bg}$. This result
coincides with the well-known result of the perfect transmission
through the waveguide bend of the type B when the rod is removed
at the corner, $\epsilon_d=\epsilon_{\rm bg}$
\cite{amjcciksfprvjdj96,sylecvhprvjdj98}.

\begin{figure}
    \centering
        \includegraphics[width=.7\columnwidth]{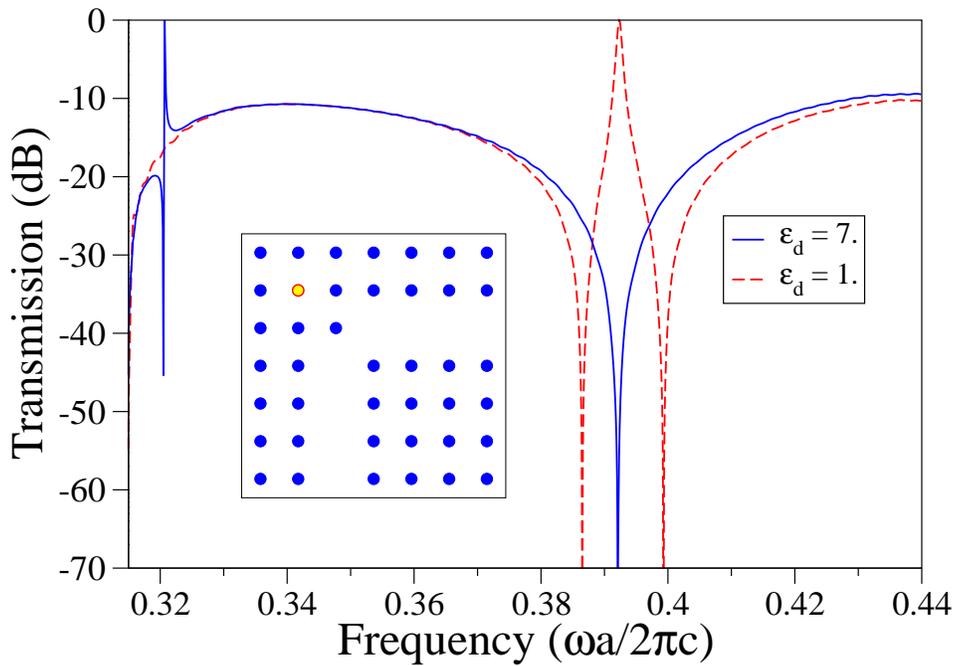}
    \caption{
Transmission coefficient through the  waveguide bend with a
(green) defect rod  placed outside the corner. In this case, there
exist two Fano resonances, one of them is characterized by an
asymmetric profile and corresponds to the perfect transmission.
     }
    \label{fig:fig4}
\end{figure}

According to Fig.~\ref{fig:fig3}, by increasing the dielectric
constant $\epsilon_d$ of the defect rod at the corner, we can
achieve the perfect resonant reflection. However, the transmission
itself becomes very low. However, it was shown in
Ref.~\cite{aemsfmsfyusk05} that when the coupling to the defect is
very small the frequency of the perfect transmission is located
very close to that of the perfect reflection; this results in a
narrow and sharp Fano resonance. We employ this idea for the
waveguide bend of the type A  and replace one lattice rod by a
defect rod outside the corner, as shown in the insert of
Fig.~\ref{fig:fig4}. In this case, we obtain the perfect
transmission for a particular frequency by varying the dielectric
constant of the defect rod, see Fig.~\ref{fig:fig4}. Here, there
exist two Fano resonances. One of them is broad, and it manifests
itself as the perfect reflection only, being similar to the case
of Fig.~\ref{fig:fig3}(c); it can be treated as the background
transmission. The second resonance possesses a sharp asymmetric
profile with both resonant transmission and reflection. Our
analysis shows that for this design the so-called \textit{double
Fano resonance} can exists due to the specific frequency
dependence of the coupling coefficients. By varying the dielectric
constant of the defect rod placed outside the corner, we shift the
asymmetric resonance whereas preserving the other one. When we
simply remove the rod outside the corner (i.e. at $\epsilon_d=\epsilon_{\rm bg}$),
these two resonances are located very close and interact with each
other. As a result, the perfect transmission is accompanied by two
perfect reflections, and this effect can be use to design a very
efficient filter based on the bend waveguide transmission.

\section{Nonlinear transmission}

Finally, we apply our effective discrete model to the case of the
nonlinear transmission through the waveguide bend with embedded
nonlinear defect rods. In this case, we assume that the (green)
defect rods placed at the bend corner (see Fig.~\ref{fig:fig2})
possesses a Kerr-type nonlinearity
\begin{eqnarray}
\epsilon_d(|E|)=\epsilon_d+\lambda|E|^2\;.
\end{eqnarray}
For definiteness, in our numerical simulations we take the value
$\epsilon_d=1.96$ and use the rescaled coefficient $\lambda=1$,
these data should correspond to some polymer materials. In the
nonlinear regime, the transmission of the waveguide bend depends
on the intensity of the incoming light. This gives us an
additional possibility to control the transmission properties of
the waveguide bend by changing the properties of the nonlinear
defect.

\begin{figure}
    \centering
        \includegraphics[width=.8\columnwidth]{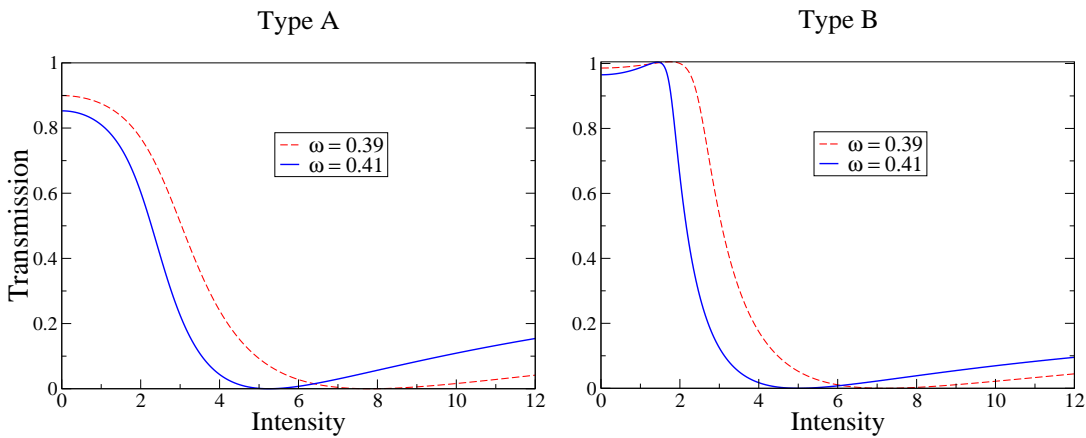}
    \caption{
Nonlinear transmission calculated for two types of the waveguide
bends shown in Fig.~\ref{fig:fig2}. In both the cases, the Fano
resonance is observed as the perfect reflection. The waveguide
bend of the type B allows the perfect transmission that can be
also tuned. }
    \label{fig:fig5}
\end{figure}

Our analysis shows that the presence of nonlinear defect does not
remove the Fano resonance itself, but instead it shifts the
position of the resonant frequency. As a result, we can tune the
value of the resonant frequency by the input light intensity, the
resonant scattering can be observed for almost all frequencies
of the waveguide spectrum~\cite{aemsfmsfyusk05}.
Figure~\ref{fig:fig5} shows the power-dependent transmission for
two types of the waveguide bends with embedded nonlinear defects
at the corner. For the type B bend,  the perfect transmission is
observed as well. The intensity-dependent transmission allows us
to control the light propagation through the waveguide bend from
$0\%$ up to $100\%$ by simply tuning the light intensity. From
another hand, we can achieve $100\%$ transmission thought the type
B bend for almost all frequencies from the spectral range by
choosing a proper intensity of the incoming light.

\section{Conclusions}

We have analyzed the conditions for high transmission through
sharp waveguide bends using the effective discrete equations
derived for two-dimensional photonic crystal waveguides. We have
demonstrated that the physics of this effect can be understood
with the help of a simple discrete model of the Fano resonance
where the waveguide bend plays a role of a specific localized
defect. Using this model, we have obtained exact analytical
solutions for the resonant transmission and reflection of two
types of the waveguide bends, in both linear and nonlinear
regimes. We believe our approach would allow to get a deeper
insight into the physics of resonant transmission through
waveguide bends, and it can be useful for understanding other
types of resonant effects in two- and three-dimensional photonic
crystal waveguides and circuits.

The authors acknowledge useful discussions and collaboration with
Sergei Mingaleev.


\end{document}